\documentclass[english]{iopart}
\usepackage{babel, mathrsfs, bbm, iopams, cite}
\usepackage{bm}
\usepackage[T1]{fontenc}
\usepackage[latin1]{inputenc}
\usepackage[bookmarks]{hyperref}

\newcommand{\mc}[1]{\mathcal{#1}}

\newcommand{\ul}[1]{\underline{#1}}
\newcommand{\Oslash}{\mc{O} \hspace{-1.4ex}/\hspace{.5ex}}
\newcommand{\Lslash}{\mbox{\L}}
\newcommand{\binom}[2]{\biggl ( \!  \! \begin{array}{c} {#1}\\{#2} \end{array} \! \! \biggr )}
\renewcommand{\exp}[1]{\mbox{exp} \left [ #1 \right ]}

\begin{document}

\title[Quantized cosm. const. in 1+1d quantum gravity]{Quantized cosmological constant in 1+1 dimensional quantum gravity with coupled scalar matter}

\author{Jan Govaerts$^{1,2,3}$ and Simone Zonetti$^{1}$}

\address{$^{1}$ Centre for Cosmology, Particle Physics and Phenomenology (CP3),
Institut de Recherche en Math\'ematique et Physique (IRMP),
Universit\'e catholique de Louvain, Chemin du Cyclotron 2, B-1348 Louvain-la-Neuve, Belgium}
\address{$^{2}$ International Chair in Mathematical Physics and Applications (ICMPA-UNESCO Chair),
University of Abomey-Calavi, 072 B. P. 50, Cotonou, Republic of Benin}
\address{$^3$ Fellow of the Stellenbosch Institute for Advanced Study (STIAS), 7600 Stellenbosch, South Africa}

\eads{\mailto{Jan.Govaerts@uclouvain.be}, \mailto{Simone.Zonetti@uclouvain.be}}

\begin{abstract} A two dimensional matter coupled model of quantum gravity is studied in the Dirac approach to constrained dynamics in the presence of a cosmological constant. It is shown that after partial fixing to the conformal gauge the requirement of a quantum realization of the conformal algebra for physical quantum states of the fields naturally constrains the cosmological constant to take values in a well determined and mostly discrete spectrum. Furthermore the contribution of the quantum fluctuations of the single dynamical degree of freedom in the gravitational sector, namely the conformal mode, to the cosmological constant is negative, in contrast to the positive contributions of the quantum fluctuations of the matter fields, possibly opening an avenue towards addressing the cosmological constant problem in a more general context.
\end{abstract}

\section{Introduction}
After close to a century since its first appearance in 1917, the cosmological constant, $\Lambda$, remains an unsolved puzzle of modern physics\cite{Weinberg:1989, Weinberg:2000, Carroll:1992, Rugh:2002}. The very first motivation for its introduction in Einstein's field equations was to counteract gravitational collapse, and ensure a static universe. Over the years observations and counter-observations made it appear and disappear, and in any case did not provide any insight into its real nature. Recent Solar System and galactic measurements and, most importantly, large scale cosmology observations (including cold dark matter models \cite{Efstathiou:1990}) have put a very stringent bound on its order of magnitude, requiring $|\Lambda| \leq 10^{-47} GeV^4$\cite{tegmark:2004}.\\
On the other hand when one accounts for the vacuum energy density in the Standard Model (SM) of particle physics the resulting value for the cosmological constant is much larger than the experimental bound. Summing up contributions from low energy QED, or the broken electroweak theory, and QCD one obtains a huge difference with observations, up to a few decades of orders of magnitude. In fact by extrapolating such estimates up to the Planck scale, without considering new physics, one gets some 120 orders of magnitude of discrepancy (see \cite{Rugh:2002} for a complete non technical overview of the contributions from the SM). It is not surprising therefore that this situation has been called ``the worst prediction in the history of physics''.\\
It has also to be considered that the vacuum energy density of the universe is influenced by phase transitions/symmetry breaking mechanisms: if one interprets the differences in energy scales between GUT, electroweak and QCD symmetry breakings as estimates for differences in vacuum energy density, namely if one accounts for the fact that the value of the cosmological constant may have changed throughout the early evolution of the Universe after the inflation era, it is likely that one should look for a coherent understanding for the value of $\Lambda$ at low energies. In particular in the present authors' view, accounting for gravity as well is a necessary first step that cannot be disregarded.\\
Gravity is the one force that drives cosmological scale dynamics, and at the same time the only (known) fundamental force that still evades even the mildest attempts of quantum unification. While it is not clear how and when a satisfactory quantum theory of gravity, or a unified theory of the fundamental interactions and their particles, will be developed, simplified scenarios of quantized matter and gravitational degrees of freedom can give interesting insight. It has been shown in a previous work (see \cite{Govaerts:2004ba} for details) that in 0+1 dimensions, when a generic matter system is coupled to a 1-dimensional metric degree of freedom and the theory is quantized, the cosmological constant, even without any dynamics in the gravitational sector, is forced to take values that belong to the matter energy spectrum, and is then itself quantized, as an effect of the time-reparametrization invariance of the model at the quantum level. Nonetheless physical quantum states are confined to the subspace of states for which the energy has the same value as the cosmological constant.\\
The idea of a quantized cosmological constant has been proposed in many different approaches: in string theory \cite{GonzalezDiaz:1987gi}, in non-commutative AdS3 \cite{Pinzul:2005ta}, in toy models of a 1-dimensional quantum gravity theory \cite{Fujikawa:1996mk} as well as in 2+1 and 3+1 dimensions in non-perturbative/loop quantum gravity \cite{Major:1995yz,Smolin:1995vq, Borissov:1996}. A link between the discrete spectrum of the volume operator and the allowed values for the cosmological constant has also been obtained in a power expansion of canonical gravity \cite{Gambini:2000ir}.
\\
This paper addresses the issue in a generalized dilaton theory (GDT) in 1+1 dimensions, by coupling Liouville gravity (see \cite{Grumiller:2002nm, Grumiller:2006rc, Nakayama:2004vk} and references therein) to an arbitrary number of real scalar fields. Gravity in two dimension has been widely studied since its first formulation by Jackiw and Teitelboim \cite{Jackiw:1982hg,Teitelboim:1983fg} and GDTs represent a very interesting class of models, which can be obtained through all kinds of compactifications from higher dimensions (for an excellent summary see \cite{Brown:1988}), like \emph{spherically reduced gravity}\cite{Berger:1972}, or dimensional reduction of pure Einstein gravity  \cite{Grumiller:2007wb}. These models are closely related to string theory with a dynamical world sheet metric \cite{Katanayev1986413, Katanaev19901}, inspiring for example the Witten black hole \cite{Witten:1991yr} and the dilaton black hole model \cite{Callan:1992zr}, and can be treated as a non-linear gauge theory \cite{Ikeda:1993aj}.
By using the BRST formalism, partially fixing the gauge symmetry, we are able to quantize the model within the Dirac procedure, obtaining a quantum realization of the constraints that eventually leads to a quantized cosmological constant.\\

The paper is organized as follows. The second Section discusses the classical theory, by moving from the Lagrangian to the Hamiltonian, and then the BRST formalisms, by fixing the gauge symmetry and calculating the gauge constraint algebra. Some classical solutions are discussed. The third Section deals with the quantum realization of the model, quantizing the ghost, the Liouville and the matter sectors. A central extension of the Virasoro algebra is identified and cancelled through an appropriate choice of quantum corrected renormalized parameters for quantum consistency of the gauge symmetries. The quantum realization of constraints on the physical states is discussed in the fourth Section, followed by a spectrum analysis that eventually leads to a quantized value of the cosmological constant inclusive of finite quantum corrections to its classical value.\\

\section{Classical theory}
\subsection{Lagrangian formulation}

From a generic two dimensional GDT model \cite{Banks:1991, Odintsov:1991} with an arbitrary number $D$ of real scalar fields, denoted by $\phi^I$, and a dynamical gravitational field, we consider an action containing a kinetic term for matter fields, their coupling to gravity and a potential term,
\begin{equation}
\mc{S} = \int d^2 x \sqrt{-g} \bigl( \mc{N} \phi_{,\alpha}^I \phi^{I ,\alpha} + \xi W( \phi^I ) R - V(\phi^I ) \bigr),
\end{equation}
where $W$ and $V$ are functions of the matter fields, the parameter $\xi$ is the coupling constant between matter and gravity, greek indices label space-time dimensions, while Einstein's convention for summing over repeated and contracted indices is implicit throughout. The usual Einstein-Hilbert action term has been omitted, since it corresponds to the Ricci scalar density $\sqrt{-g}R$ which is a total derivative in two dimensions. Note that in the absence of the coupling $\xi$, the gravitational sector is thus not dynamical and decouples altogether, a situation specific to two dimensions.\\
Taking for the metric the $(-,+)$ (Lorenzian) signature, and choosing the canonical normalization for scalar fields, we set $\mc{N}=-1/2$.\\
The topology of the 1+1 dimensional spacetime is taken to be that of a cylinder, namely $\mc{M} = \mathbb{R} \times S$, so that any space dependence is periodic for the coordinate interval $\sigma \in [0, 2\pi)$, while the time coordinate is denoted as $\tau$.
The metric tensor may be parametrized with 3 independent fields, $\varphi$ and $\lambda^\pm$, 
so to have a line elements which reads:
\begin{equation}\label{metric}
ds^2=e^{\varphi}\left ( -\lambda^- \lambda^+d\tau^2+(\lambda^+ -\lambda^-)d\tau\ d\sigma+d\sigma^2\right )
\end{equation}
The Ricci scalar density then reduces to the form $\sqrt{-g}R = M(\lambda^+, \lambda^-, \varphi)_{,\sigma} + S(\lambda^+, \lambda^-, \varphi)_{,\tau}$, where $M$ and $S$ are known functions. From now on without any risk of ambiguity we drop commas denoting derivatives with respect to $\tau$ and $\sigma$. Through integration by parts, one thus has,
\begin{equation}\label{actionMS}
\mc{S} = \int d \tau d \sigma  \left [ \sqrt{-g}\left(-\frac{1}{2} \phi^I_{\alpha} \phi^{I \alpha} - V(\phi^I) \right ) - \xi  (W_\tau M + W_\sigma S) \right ].
\end{equation}
Henceforth only linear and identical couplings between the Ricci scalar $R$ and the scalar fields will be considered, so that $W(\phi^I) = \sum \phi^I$. Furthermore the potential term is restricted to being constant, $V(\phi^I)=\Lambda$, corresponding then to a classical cosmological constant term.

\paragraph{Equivalence to Liouville gravity.} It is quite straightforward to show that such a model is in fact equivalent to Liouville gravity, with a specific choice of the parameters, and $D-1$ minimally coupled scalars. Defining:
\begin{equation}\label{overallscale} \eqalign{
X:=\xi\ \sum_{I=1}^{D}\phi^{I},\\
\theta_{i} :=  \frac{\xi \sqrt{D}}{ \sqrt{ i(i+1)}} \left [ \sum_{I=1}^i \phi^I - i \phi^{i+1} \right ] \qquad n = 1, \ldots, D-1,}
\end{equation}
and rescaling the conformal mode and the cosmological constant:
\begin{equation}
\varphi\rightarrow\tilde{\varphi} D^{-1}\xi^{-2} \qquad \Lambda \rightarrow \tilde{\Lambda} D^{-1}\xi^{-2}
\end{equation}
the action \eref{actionMS} can be rewritten as:
\begin{equation}\label{actionliouville}
\mathcal{S}=\frac{1}{D\xi^2}\int d^{2}x\sqrt{-\tilde{g}}\left(-\frac{1}{2}\sum_{i=1}^{D-1}\theta_{i,\alpha}\theta_i^{,\alpha}-\frac{1}{2}X_{,\alpha}X^{,\alpha}+X\ \tilde{R}-\tilde{\Lambda}\right)
\end{equation}
where $D\xi^2$ becomes an overall scale. This is the Liouville gravity action presented in \cite{Grumiller:2006rc} with $\alpha=0, a=-\frac{1}{2}, b=\frac{\tilde{\Lambda}}{2}$. 

\paragraph{Field redefinitions.}
It is however more convenient in the quantization procedure to keep $\xi$ as a coupling constant, which will acquire quantum corrections.
For this reason one can consider the fields:
\begin{equation}\label{decoupling} \eqalign{
X:=\xi\ \sum_{I=1}^{D}\phi^{I},\\
B_{i} :=  \frac{1}{\sqrt{i(i+1)}} \left [ \sum_{I=1}^i \phi^I - i \phi^{i+1} \right ] \qquad n = 1, \ldots, D-1,}
\end{equation}
and, following \cite{LouisMartinez:1993eh}, a conformal transformation can be applied to the metric, in the form $\varphi \rightarrow \varphi + \frac{X}{D \xi^2}$. Then, by defining the new fields  $A:=\sqrt{D}\xi\varphi + X$ and $B_\emptyset:=\varphi \xi \sqrt{D} $ all degrees of freedom decouple from each other. Writing the metric tensor as in \eref{metric}, the lagrangian density finally takes the form:
\begin{equation}\label{decoupledL}
\eqalign{
\mc{L} &= -\sum_{i=1}^{D-1} \mc{L}(B_i) - \mc{L}(B_\emptyset) + \mc{L}(A)+\frac{1}{2} \Lambda e^{\frac{A }{\sqrt{D}\xi }} (\lambda^+ +\lambda^- )+\\
&+ \frac{2 \sqrt{D} \xi }{\lambda^+ + \lambda^-} \left ( \bigl ( A - B_\emptyset \bigr )_\tau \bigl ( \lambda^+ - \lambda^-\bigr )_\sigma + 2 \bigl ( A - B_\emptyset \bigr )_\sigma \bigl ( \lambda^+ \lambda^-\bigr )_\sigma\right ) ,
}
\end{equation}
where for any of the fields, $f$, the expression $\mc{L}(f)$ stands for the expression 
\begin{equation}
\mc{L}(f) = \frac{\left( \lambda^+  f_{\sigma }-f_{\tau } \right ) \left( \lambda^-  f_{\sigma }+f_{\tau }\right )}{\lambda^+ +\lambda^-}.
\end{equation}

\subsection{Classical solutions}
It is worth to briefly consider some of the classical solutions of such model.
\paragraph{General solutions.} In the following the \emph{conformal gauge} will be adopted, with the conditions $\lambda^+ = \lambda^- = \lambda_0$, $\lambda_0$ being an arbitrary positive constant, $\lambda_0>0$. These conditions will be also chosen in the quantum theory.
In this way from \eref{decoupledL}  one obtains for the equations of motion and constraints:
\numparts
\begin{eqnarray}\label{classicaleoms}
\lambda_0^2 B_{i, \sigma \sigma} - B_{i, \tau \tau} = 0,\qquad i = \emptyset, 1, \ldots D-1,\label{kgeom}\\
\lambda_0^2 \left(\frac{\Lambda}{\sqrt{D}\xi}\exp{\frac{A}{\sqrt{D}\xi}}+A_{\sigma \sigma} \right ) - A_{\tau \tau}=0,\label{liouvilleeom}\\
\eqalign{
& -\left [ \bigl( \partial_\tau \pm \lambda_0 \partial_\sigma \bigr) A \right ]^2 +  \left [ \bigl( \partial_\tau \pm \lambda_0 \partial_\sigma \bigr) B_\emptyset \right ]^2 + \\ & \quad +\sum_{i=1}^{D-1} \left [ \bigl( \partial_\tau \pm \lambda_0 \partial_\sigma \bigr) B_i \right ]^2  +2 \sqrt{D} \xi \bigl( \partial_\tau \pm \lambda_0 \partial_\sigma \bigr)^2 (A- B_\emptyset) = 0,
\label{classc}
}
\end{eqnarray}
\endnumparts
where again subscripts $\tau$ and $\sigma$ denote partial derivatives.

Equations \eref{kgeom} identify $D$ massless scalar fields, while \eref{liouvilleeom}, for $\Lambda \neq 0$, is the typical Liouville field equation, with the general space and time dependent solution:
\begin{equation}\label{afieldsol}
A(\tau, \sigma) = \sqrt{D} \xi \ ln \! \left [ \frac{8 D \xi^2}{\Lambda} \frac{\partial_+ u^+ (x^+) \partial_- u^-(x^-)}{\left ( u^+ (x^+) - u^- (x^-)\right )^2}\right ],
\end{equation}
where $u^\pm$ are two arbitrary functions of $x^\pm = \lambda_0 \tau \pm  \sigma$, respectively, and $\partial_\pm$ is the derivative with respect to $x^\pm$. When $\Lambda = 0$ \eref{liouvilleeom} reduces to the massless Klein-Gordon equation.\\
The two remaining equations further restrict the classical fields: both of them are independent from the value of the cosmological constant $\Lambda$, which then appears only in \eref{afieldsol}. This is easily seen by expanding the $B$ fields as $B_i(\tau, \sigma) = B_i^+(x^+) + B_i^-(x^-)$ and using \eref{afieldsol}, in the case $\Lambda\neq0$. Then \eref{classc} can be rewritten in the form:
\begin{equation}\label{pmconstraints}
\eqalign{
&D \xi ^2 \left [ 3\left( \partial_\pm ^2 u^\pm \right  )^2-2 \partial_\pm u^\pm \partial^3_\pm u^\pm\right ]-\\&-\left (\partial_\pm u^\pm\right  )^2 \left [2 \sqrt{D} \xi \partial_\pm^2 B_\emptyset^\pm - \left (\partial_\pm B_\emptyset^\pm\right )^2+ \sum_{i=1}^{D-1} \left(\partial_\pm B_i^\pm\right )^2\right]=0.
}
\end{equation}
Since $\Lambda$ does not appear in \eref{pmconstraints} we can conclude that the only classical restriction on the cosmological constant, namely on its sign, is to ensure the existence of the logarithm \eref{afieldsol}. The classical cosmological constant is then a simple parameter in the classical solutions.\\
Moreover the model is equivalent to two systems of arbitrary right- and left-moving scalar fields which satisfy \eref{pmconstraints}, further constrained by spatial periodicity. It is however quite complicated to reconcile both requirements, mainly because of the logarithmic form of \eref{afieldsol} and obtain explicit solutions.\\
It is worth noting that $A=const$ is not a solution of \eref{liouvilleeom}, unless $\Lambda = 0$. Hence the classical ground state, {\it i.e.} when all fields vanish (up to constant shifts) and space-time reduces to a rescaled Minkowski solution, only allows for a vanishing cosmological constant.\\
The Ricci scalar is easily calculated as:
\begin{equation}
R=\frac{e^{-\frac{A}{\sqrt{D}\xi}}}{\sqrt{D}\xi} \left(\lambda_{0}^{-2}\partial_{\tau}^{2}-\partial_{\sigma}^{2}\right)A=\frac{\Lambda}{D\xi^2},
\end{equation}
so that space-time curvature is constant and fully determined by $\Lambda$ and the choice of parameters, with no dependence on the gravitational degrees of freedom. This is a peculiar feature of Liouville gravity.

\paragraph{The homogeneous case.} As an example, a very simple case can be studied, looking for homogeneous, {\it i.e.}, space-independent, solutions. All the $B$ fields are just linear functions of time, in the form:
\begin{equation}
B_i = b^{(0)}_i + b^{(1)}_i\, \lambda_0 \tau, \qquad i = \emptyset, 1, \ldots, D-1,
\end{equation}
so that the constraint equations reduce to:
\begin{equation}\label{homeoms}
A_\tau^2-2 \sqrt{D} \xi A_{\tau \tau} = \mc{B}^2, \quad \mbox{with} \quad \mc{B}^2 =  (b^{(1)}_\emptyset)^2 +  \sum_{i=1}^{D-1} (b^{(1)}_i)^2.
\end{equation}
For a non-vanishing $\mc{B}>0$ with $\Lambda\neq0$, only hyperbolic solutions are admitted, depending on the sign of $\Lambda$. So one has:
\begin{equation}\label{homogenouscase}
\exp{A(\tau)/ \sqrt{D}\xi} = \left \{ \begin{array}{l}
\frac{ \mc{B}^2}{2|\Lambda| \lambda_0^2}  \sinh^{-2}\! \left( \frac{\mc{B}}{2 \sqrt{D} \xi}(\tau - \tau_0) \right)\ : \quad  \Lambda>0,\\
\frac{ \mc{B}^2}{2|\Lambda| \lambda_0^2} \cosh^{-2} \! \left(\frac{\mc{B}}{2 \sqrt{D} \xi}(\tau - \tau_0)\right)\ : \quad  \Lambda<0,
\end{array} \right.
\end{equation}
which is regular in the $\Lambda<0$ case, with a closed universe displaying an accelerating expansion followed by a decelerating expansion and a contraction. 
For $\mc{B} = 0$ on the other hand one gets a singular solution:
\begin{equation}
\exp{A(\tau)/ \sqrt{D}\xi}  = \frac{2 D \xi^2}{\Lambda \lambda_0^2 (\tau - \tau_0)^2},
\end{equation}
with the requirement $\Lambda>0$.\\
A vanishing cosmological constant would further simplify the equations of motion for $A$, which would reduce to $A(\tau) = A_0 \pm \mc{B}\lambda_0(\tau-\tau_0)$, allowing for an expanding, contracting or static universe.

\paragraph{Vacuum Solutions.} Starting from the dilaton action \eref{actionliouville} it is easy to find a general solution in the case of constant matter fields, \emph{i.e.} with $\theta_i=\mbox{const}$, following the results of \cite{Grumiller:2002nm}.\\
Using the Eddington-Finkelstein gauge, and considering only the region $\tau>0$, the line element is $ds^{2}=2d\tau d\sigma+2K(\tau,\sigma)\ d\sigma^{2}$, and the equations of motion are solved by:
\begin{eqnarray*}
X=&-2\ln\left(\tau\ x_{0}(\sigma)\right)\\
K=&\frac{\tilde{\Lambda}}{4}\tau^{2}+k_{1}(\sigma)\tau+k_{0}(\sigma)
\end{eqnarray*} 
where $x_{0},k_{1},k_{0}$ are arbitrary periodic functions of $\sigma$, with the requirement of $X$ to be well-defined.
The two remaining equations are constraints which can be solved for the $k$'s, so that:
\begin{equation}
K(\tau,\sigma)=\frac{\tilde{\Lambda}}{4}\tau^{2}\left(1+4\frac{\alpha+x_{0}'(\sigma)}{\tilde{\Lambda}x_{0}(\sigma)}\tau^{-1}\right) 
\end{equation}
where $\alpha$ is an arbitrary constant. There will then be an horizon at $\tau=-4\frac{\alpha+x_{0}'(\sigma)}{\tilde{\Lambda}x_{0}(\sigma)}$, where $K=0$.
It is also worth pointing out that for $\tau\rightarrow\infty$ space-time is asymptotically AdS. Such a feature is present also in other classical solutions in the presence of matter.\smallbreak
A more detailed analysis of these interesting features of the classical solution, being beyond the scope of this paper, will be developed in a separate work.

\subsection{Hamiltonian Formulation}
With the new definition of the fields the Hamiltonian density reads,
\begin{equation}\label{hamiltoniandensity}
\eqalign{
\fl \mc{H} & = \lambda^-  \left[ \left ( \frac{1}{2} e^{\frac{A }{ \sqrt{D} \xi }} \Lambda-\frac{1}{4}\left(\chi_A^+\right)^2 +  \sqrt{D} \xi   \chi_{A\sigma}^{+} \right ) + \left ( \frac{1}{4}  \left( \chi_\emptyset^- \right)^2 - \sqrt{D} \xi  \chi_{\emptyset \sigma}^- \right ) + \sum_{i=1}^{D-1}\frac{1}{4}\left(\chi_i^-\right)^2 \right ]+ \\
\fl & +\lambda^+ \left [ \left ( \frac{1}{2} e^{\frac{A }{ \sqrt{D} \xi }} \Lambda-\frac{1}{4}\left(\chi_A^-\right)^2 - \sqrt{D} \xi  \chi_{A\sigma}^{-} \right ) + \left ( \frac{1}{4}  \left( \chi_\emptyset^+ \right)^2 + \sqrt{D} \xi  \chi_{\emptyset \sigma} ^+ \right ) + \sum_{i=1}^{D-1} \frac{1}{4} \left(\chi_i^+\right)^2 \right ],
}
\end{equation}
where
\begin{equation}
\chi_A^\pm = \Pi_A \pm A_\sigma, \qquad \chi_\emptyset^\pm = \Pi_\emptyset \pm B_{\emptyset\sigma}, \qquad \chi_i^\pm = \Pi_i \pm B_{i\sigma},
\end{equation}
while the $\Pi$'s are the canonically conjugate momenta of the fields (the subscript $\sigma$ denotes a derivative with respect to the space coordinate). The coordinate dependence is implicit for all the quantities involved. \\
While the conjugate momenta of the $A$ and $B$ fields are well defined, one has $\Pi_{\lambda^\pm}=0$, showing that the system possesses two primary constraints, $L^1=\Pi_{\lambda^+}$ and $L^2=\Pi_{\lambda^-}$, as it should given its two dimensional reparametrization invariance. The consistency conditions under dynamical evolution for the (smeared) primary constraints, $\dot{L}^\mu = \{ L^\mu , H \} = 0$, lead to two further secondary constraints, denoted $L^+$ and $L^-$. These constraints correspond to the two quantities in square brackets being multiplied by the $\lambda^+$ and $\lambda^-$ fields in \eref{hamiltoniandensity}, respectively. Given the decoupling achieved through the field redefinitions in \eref{decoupling} these secondary constraints are simply expressed as a sum of separate terms, one for each different field,
\begin{equation}\label{matterconstraints}
L^\pm = L^{\pm,A}+ L^{\pm,\emptyset} + \sum_{i=1}^{D-1} L^{\pm,i}.
\end{equation}
The Hamiltonian density itself is a linear combination of these constraints, and is therefore weakly vanishing as expected. In turn, these secondary constraints have to fulfil once again the same requirement of a consistent time evolution, a criterion which is readily ascertained once their algebra of Poisson brackets is known. In the decoupled field basis being used, the (smeared) algebra of these constraints in the $B_i$ sectors factorizes, leading to,
\numparts
\begin{eqnarray}\label{sectorsalgebra}
\{ L^{\pm,{A+\emptyset}} (f) , L^{\pm,{A+\emptyset}} (g) \} & = \pm L^{\pm,{A+\emptyset}} (f g_\sigma - f_\sigma g), \\
\{ L^{+,{A+\emptyset}} (f) , L^{-,{A+\emptyset}} (g) \} &= 0,\\
\{ L^{\pm,i} (f) , L^{\pm,i} (g) \} &= \pm L^{\pm,i} (f g_\sigma - f_\sigma g), \\
\{ L^{-,i} (f) , L^{+,i}(g) \} &= 0,
\end{eqnarray}
\endnumparts
where an integration over the space coordinate $\sigma$ is always implied for the smearing test functions $f$ and $g$ which multiply the constraints.\\
The closure of this algebra ensures that the consistency condition is always satisfied for a system with the $A$, $B_\emptyset$ and an arbitrary number of $B_i$ fields. It is worth noting that the $A$ and $B_\emptyset$ sectors do not exhibit a closed algebra if taken separately, because of their coupling to the gravitational field variables $\lambda^\pm$. However, such non vanishing contributions cancel one another when the two sectors are taken in combination.

\paragraph{BRST Formulation.}
Given the set of four constraints $L^1$, $L^2$, $L^- $ and $L^+$ one may introduce four pairs of anticommuting canonically conjugate \emph{BRST ghosts} $c^a, \mc{P}_a$, with $a$ taking the values $a=1,2,-,+$, and with ghost numbers $\pm1$, respectively \cite{Govaerts:1991}. In contradistinction to this \emph{ghost sector}, the set of the $B$ and $A$ fields will be referred to as the \emph{bosonic sector}.
The BRST charge $Q_B$ is defined to be real, of ghost number $g_c = 1$, Grassmanian odd, nilpotent and such that $\frac{\partial}{\partial c^a} Q_B | _{c^a = \mc{P}_a = 0} = L^a$.
One can easily see that the expression:
\begin{equation}\label{QBRST}
Q_B = \int d \sigma \left ( c^a L^a - c^- c^-_\sigma \mc{P}_- + c^+ c^+_\sigma \mc{P}_+ \right),
\end{equation}
meets all these properties.
Through Poisson brackets, the action of the BRST charge on the ghost variables $\mc{P}_-$ and $\mc{P}_+$ gives the BRST extension of the constraints $L^-$ and $L^+$,
\begin{equation}
\delta_{\rm brst} \mc{P}_\pm = - \left( L^\pm \pm 2 c^\pm_\sigma \mc{P}_\pm \pm c^\pm \mc{P}_{\pm \sigma} \right ) = -L^{\pm,{brst}}.
\end{equation}
One can directly check that both expressions in brackets, $L^{\pm,{brst}}$, fulfil the requirements for BRST extended observables, and exhibit the same algebra as the original constraints constructed out of the original fields only, $L^\pm$.\\
The BRST extensions of the Hamiltonian density can be obtained using an arbitrary function $\Psi$ on extended phase space, of odd Grassmann parity, of ghost number $g_c = -1$, and which is anti-hermitian. The complete BRST Hamiltonian density then reads,
\begin{equation}\label{BRSThamiltonian}
\mc{H}_{brst}=  - \{ \Psi, Q_B \}.
\end{equation}
Suitable boundary conditions have to be considered for the ghost sector. These extended degrees of freedom need to be periodic in the $\sigma$ spatial coordinate, and to be vanishing at $\tau$-infinity.

\paragraph{Gauge Fixing.}
In order to proceed to the quantization of the theory we partially fix the gauge freedom in spacetime diffeomorphisms which is parametrized by the Lagrange multipliers $\lambda^\pm$, and yet carefully avoid Gribov problems \cite{Govaerts:1991, Gribov:1977wm}. Choosing to work in the \emph{conformal gauge}, we can fix a specific form for the $\Psi$ function \cite{Govaerts:1991},
\begin{equation}
\Psi = \frac{1}{\beta} \mc{P}_1 (\lambda^+ - \lambda_0) +  \frac{1}{\beta} \mc{P}_2 (\lambda^- - \lambda_0) +  \mc{P}_-\lambda^+ +  \mc{P}_+\lambda^-,
\end{equation}
where $\beta$ is a free real parameter that will be taken to vanish later on. By computing the equations of motion for the phase space variables generated by the BRST extended Hamiltonian \eref{BRSThamiltonian}, then rescaling the fields in order to absorb the factor $\beta$,
\begin{equation}
\begin{array}{c c c c}
\mc{P}_1\Rightarrow \frac{i}{2 \pi} b_- \beta, & \mc{P}_2\Rightarrow \frac{i}{2 \pi} b_+ \beta, & \Pi_{\lambda^-} \Rightarrow \tilde{\Pi}_{\lambda^-} \beta ,& \Pi_{\lambda^+} \Rightarrow \tilde{\Pi}_{\lambda^+} \beta,
\end{array}
\end{equation}
and then taking the limit $\beta \rightarrow 0$, one obtains,
\begin{equation}
\begin{array}{c c}
b_- =  2 \pi i \mc{P}_-, & b_+ =  2 \pi i \mc{P}_+, \\
\tilde{\Pi}_{\lambda^+} = - L^-, & \tilde{\Pi}_{\lambda^-} = - L^+, \\
\lambda^- = \lambda^+ = \lambda_0, & c^1 = c^2 = 0.
\end{array}
\end{equation}
The quantity $\lambda_0$, which parametrizes the different gauge orbits, can always be set to unity by using the residual gauge freedom $\tau \rightarrow \lambda_0 \tau$, is left explicit for clarity.\\
By imposing these on-shell conditions on the BRST Hamiltonian density \eref{BRSThamiltonian} one gets,
\begin{eqnarray*}
\mc{H} &=& \lambda_0 \left ( L^{-,brst} + L^{+,brst} \right ), \\
L^{\pm,brst} &=& L^\pm \mp \frac{i}{2\pi}\left(2c^\pm_\sigma b_\pm + c^\pm b_{\pm\sigma}\right)=:L^\pm+L^{\pm,g}
\end{eqnarray*}
while the BRST charge \eref{QBRST} becomes,
\begin{equation}
Q_B = \int d \sigma \left ( c^- L^- + c^+ L^+ +\frac{i}{2 \pi} \left ( c^- c^-_\sigma  b_- - c^+ c^+_\sigma b_+\right ) \right ).
\end{equation}
Since the ghosts $(c^\pm,b_\pm)$ are canonically conjugate, $\{c^\pm(\sigma),b_\pm(\sigma')\}=-2i\pi\delta_{2\pi}(\sigma - \sigma')$ (with $\delta_{2\pi}(\sigma-\sigma')$ being the $2\pi$-periodic Dirac $\delta$ distribution on the unit circle), one may check that $Q_B$ is still nilpotent, {\it i.e.}, it has a vanishing Poisson bracket with itself.

\paragraph{The constraint algebra.}
The two BRST extended constraints obey the smeared algebra:
\begin{equation}
\{ L^{\pm,brst} (f) , L^{\pm,brst} (g) \} = \pm L^{\pm,brst} (g_\sigma f- f_\sigma g),
\end{equation}
hence their equations of motion, computed with the gauge fixed Hamiltonian, are:
\begin{equation}
L^{\pm,brst}_\tau = \pm \lambda_0 L^{\pm,brst}_\sigma,
\end{equation}
which admit as solutions the mode expansions:
\begin{equation}\label{constraintsexpansion}
L^{\pm,brst} = \sum_{n \in \mathbb{Z}} \frac{1}{2\pi}L^{\pm,brst}_n \exp{-i n (\lambda_0 \tau \pm \sigma)}.
\end{equation}
In this way it is straightforward to compute the algebra for the modes $L^{\pm,brst}_n$ through a Fourier transformation,
\begin{equation}
\{ L^{\pm,brst}_n, L^{\pm,brst}_m \} = - i (n-m) L^{\pm,brst}_{n+m}.
\end{equation}
For each chiral sector this is the celebrated Virasoro algebra, {\it i.e.}, the partially gauge fixed classical theory is a \emph{conformal invariant theory}. In particular, note how the two chiral sectors do commute with one another.

\section{Quantum theory}

Quantization follows the usual canonical procedure, by choosing a polarization for phase space and a Hilbert space, by promoting observables to operators (using normal ordering when required, with the annihilation operators to the right of the creation ones) and by substituting Poisson brackets with commutators or anticommutators, as the case may be, inclusive of the extra factor $i\hbar$ multiplying the values of the corresponding classical brackets.

\subsection{Quantization and Constraint Algebra}

\paragraph{The ghost sector.} In the ghost sector, using a Fock basis to span Hilbert space, and since the classical ghost fields $c^\pm$ and $b_\pm$ are Grassmann odd variables, the quantized theory obeys the Fermi-Dirac statistics.\\
For the ghost operators, omitting the hat emphasizing the operator character of observables, one defines the following mode expansions, at time $\tau = 0$,
\begin{equation}\label{ghostsexpansions}
c^\pm (\sigma) = \sum_{n \in \mathbb{Z}} c^\pm _n \exp{\mp i n \sigma}, \qquad 
b_\pm (\sigma) = \sum_{n \in \mathbb{Z}} b_\pm ^n \exp{\mp i n \sigma},
\end{equation}
with $\{c^\pm_n,b^\pm_m\}=\hbar \delta_{n+m,0}$, ${c^\pm_n}^\dagger=c^\pm_{-n}$ and ${b^n_\pm}^\dagger=b_\pm^{-n}$.
The vacuum of the theory, denoted as $|\Omega \rangle$, is the tensor product of all vacua for all modes $n\in\mathbb{Z}$, which are annihilated by all the $c^\pm_n$'s, are normalized to 1, and are such that all $b^n_\pm$'s act as creation operators. Therefore one has a countable infinity of ghost/anti-ghost pairs of operators, one such pair for each $n\in\mathbb{Z}$ in \eref{ghostsexpansions}.\\
Given the decoupling of the two chiral sectors of the ghost variables in the conformal gauge, both for the canonically conjugate pairs of ghost degrees of freedom as well as their contributions to the constraints, an efficient way to compute the quantum ghost Virasoro algebra is through \emph{radial quantization} \cite{Difrancesco:1999, Ginsparg:1988ui, Blumenhagen:2009zz}. The quantum algebra for the modes of the ghost contributions to the total Virasoro generators is then found to read:
\numparts
\begin{eqnarray}\label{ghostalgebra}
\bigl[\hat{L}^{\pm,g}_n,  \hat{L}^{\pm,g}_m \bigr] &= (n-m)\hbar \hat{L}^{\pm,g}_{m+n} -\hbar^2\frac{13}{6}(n^3 - n) \delta_{m+n},\\
\bigl[\hat{L}^{+,g}_n,  \hat{L}^{-,g}_m \bigr] &= 0.
\end{eqnarray}
\endnumparts
Hence, as it is well known, the Virasoro algebra in the ghost sector acquires a quantum central extension, namely a quantum anomaly which breaks the conformal symmetry of the classical ghost sector.

\paragraph{The bosonic sector.}
Starting from the classical constraints, $L^\pm$, contributing in \eref{hamiltoniandensity}, in order to ensure the closure of their quantum algebra and the cancellation of the total quantum central charges of the total Virasoro generators, the possibility of quantum corrections to the coupling constant $\xi$ needs to be considered \cite{Curtright:1982}, in a manner dependent on the fields. As a matter of fact, only terms involving the fields linearly need to be corrected, namely the $A$ and $B_\emptyset$ fields only with the replacements $\xi \rightarrow \xi_A = \xi + \delta_A$ and $\xi \rightarrow \xi_\emptyset = \xi + \delta_\emptyset$ for the corresponding couplings, respectively. The factor $\xi$ appearing in the exponential Liouville term contribution to $L^\pm$ remains unchanged,
\begin{equation}\label{qvirasoro}
\eqalign{
L^\pm(\sigma) =&  \left [ \frac{1}{2}\Lambda\, e^{\frac{A }{\sqrt{D} \xi }} -\frac{1}{4}\left(\chi_A^\mp\right)^2 \mp \sqrt{D} \xi_A  \left( \chi_A^{\mp}\right)_\sigma \right ]+\\&+ \left [ \frac{1}{4}  \left( \chi_\emptyset^\pm \right)^2 \pm \sqrt{D} \xi_\emptyset  \left(\chi_\emptyset^\pm \right)_\sigma \right ]+\frac{1}{4} \sum_{i=1}^{D-1}\left(\chi_i^\pm\right)^2.
}
\end{equation}
As may be seen from \eref{hamiltoniandensity} and \eref{matterconstraints}, terms associated to different fields have a similar form. Hence the computation of the quantum algebra for the $L^{\pm,A}$ contributions provides the general result which may be particularized to all other fields. However, because of the Liouville exponential term involving the $A$ field, radial quantization can no longer be used: even if $A$ is expressed as the sum of holomorphic and antiholomorphic contributions, the exponential coupling between these two sectors through the Liouville potential does not allow to separate the two complex variables. Consequently one has to consider Fourier mode expansions of the fields and compute directly the commutators for these modes.\\
Following \cite{Curtright:1982} the field $A$ and its conjugate momentum $\Pi_A$ are expressed in terms of a creation/annihilation zero-mode pair $(a_0, a^\dagger_0)$ and two chiral sets of  non-zero mode Fock operators, $a_n, \bar{a}_n$ for $n \neq 0$, $n\in\mathbb{Z}$, where positive (resp., negative) $n$'s correspond to annihilation (resp., creation) operators. Given the singularities that arise from local products of operators at the same spatial point, a regularization procedure needs to be introduced to define infinite sums over field modes. For convenience of computation, we have opted for a simple exponential damping regularization factor $e^{-\varepsilon |n|}$, with $\varepsilon\rightarrow 0^+$, to be included in all field mode expansions,
\numparts
\begin{eqnarray}\label{Amodes}
A(\sigma) &= \frac{i}{2 \sqrt{\pi} } \left [ a_0 - a^\dagger_0 + {\sum_n}' \frac{1}{n}\left ( a_n e^{-in\sigma} + \bar{a}_ne^{in\sigma} \right ) e^{-\varepsilon |n|} \right ],\\
\Pi_A(\sigma) &= \frac{1}{2 \sqrt{\pi}} \left [ a_0 + a^\dagger_0 + {\sum_n}' \left ( a_n e^{-in\sigma} + \bar{a}_ne^{in\sigma} \right ) e^{-\varepsilon |n|} \right ],
\end{eqnarray}
\endnumparts
where the primed sum, $\sum'_n$, stands for a sum over all non zero modes, $n\ne 0$, $n\in\mathbb{Z}$. The given mode operators obey the following algebra of commutation relations,
\begin{equation}
[a_n , a_m ] = [\bar{a}_n, \bar{a}_m ] = n\hbar \delta^n_{-m}, \ a^\dagger_n=a_{-n},\
\bar{a}^\dagger_n=\bar{a}_{-n}, \ [a_0, a^\dagger_0] = \hbar.
\end{equation}
In terms of the fields $A$ and $\Pi_A$, these commutation relations translate to the required Heisenberg algebra, once the
limit $\varepsilon\rightarrow 0^+$ is applied.

We can then carefully evaluate the commutators between the $L^{\pm,A}$, then take the limit $\varepsilon \rightarrow 0$ and choose the appropriate quantum correction for the coupling constant $\xi_A$ as $ \delta_A=\frac{\hbar}{8 \sqrt{D} \pi \xi}$, obtaining again a quantum Virasoro algebra with a central extension. This calculation, as said before, provides also the algebra for $L^{\pm,\emptyset}$\footnote{This algebra does not need a quantum correction of $\xi_\emptyset$ to close. In fact, as said before, this correction will be fixed later to cancel part of the central charge.} and for $L^{\pm,i}$. Each sector will contribute to the central charge. The interested reader can find details of the computation in \ref{matterquantization}.
\smallbreak
Putting together all contributions, from the ghost and bosonic sectors, we get for the modes of the total BRST extended Virasoro generators:
\begin{equation}\label{quantumvirasoroextended}
\eqalign{
\left [ L_n^{\pm,brst} , L_m^{\pm,brst} \right ] = (n-m) \hbar L^{\pm,brst}_{n+m} -\hbar^2\left [ \frac{D-1}{12} -\frac{13}{6} \right ] n \delta_{n+m} -\\-  \hbar \left\{ 4 D \pi \biggl [  \left ( \xi - \frac{\hbar}{8 \sqrt{D} \pi \xi} \right )^2 - \xi^2_\emptyset \biggr ] + \hbar \left (\frac{D-1}{24} +\frac{13}{6}\right )\right \} n^3 \delta_{n+m}.
}
\end{equation}
Using the freedom to fix the quantum correction in $\xi_\emptyset = \xi + \delta_\emptyset$ we may cancel the $n^3$ term in the central extension with the choice:
\begin{equation}
\delta_\emptyset =-\xi+\mbox{sgn}(\xi)\sqrt{\xi ^2+\frac{\hbar }{96 \pi }+\frac{17 \hbar }{32 D \pi }-\frac{\hbar }{4 \sqrt{D} \pi }+\frac{\hbar ^2}{64 D \pi ^2 \xi ^2}}
\end{equation}
a quantity which is real for any real values for $\xi$ and $D$ and recovers the right coupling in the classical limit. Using that freedom we are thus left with a central charge that affects only the zero modes of the $L$'s, which may be redefined as $L^{\pm,brst}_0 \Rightarrow L^{\pm,brst}_0 - \hbar^2(D - 27)/24$\footnote{There is no contradiction with \cite{Kummer:1996hy}, where it is shown that dilaton gravity in absence of matter is quantum trivial, {\it i.e.} the quantum effective action is equivalent to the classical one. The action quantized in this work is obtained after a conformal transformation and field redefinitions that change the global structure of the theory already at the classical level, inducing possible global quantum effects\cite{Katanaev:1996ni}.}, hence giving an algebra which is finally free of central extensions:
\begin{equation}\label{quantumvirasoro}
\left [ L_n^{\pm,brst} , L_m^{\pm,brst} \right ] = (n-m)\hbar L^{\pm,brst}_{n+m}.
\end{equation}
One can easily check that this gives also a nilpotent BRST charge.

\subsection{Quantum Constraints}
Once the quantum Virasoro algebra is obtained, it is possible to find the quantum realization of the constraints on Hilbert space, following the usual Dirac prescription that physical states have to be annihilated by the constraints.
As a matter of fact the cosmological constant $\Lambda$ (and the coupling constant $\xi$) are still free parameters: by requiring certain quantum states to be physical, {\it e.g.}, the Fock vacuum, $\Lambda$ will be constrained to take a specific value.\\
Taking advantage of the BRST invariance of the model, the ghost sector contributions to the operators will be dropped.\\
Again the Liouville potential involving the $A$ field prevents one from following the most direct approach, {\it i.e.}, extracting the modes $L_n^\pm$ of the quantum constraints with a discrete Fourier transform and looking for the states that satisfy $L^\pm_n |\psi \rangle=0$ with $n = 0,1,2,\ldots$, as in ordinary String Theory \cite{Green:1987sp, Polchinski:1998rq}. For our purpose, however, it is sufficient to use the weaker condition:
\begin{equation}\label{quantumconstraints}
\langle \psi | L^\pm(\sigma) | \psi \rangle = 0,
\end{equation}
in the hypothesis that $|\psi\rangle$ is physical. 
The $\sigma$ dependence will have to be carried through and in some cases traded for a mode expansion {\it via} a Fourier transformation once the matrix elements between suitable states spanning the Hilbert space have been calculated. In particular, considering linear combinations of the shifted Virasoro generators, the quantum constraints for an arbitrary quantum physical state will be:
\numparts
\begin{equation}\fl
\eqalign{
\langle L^+ + L^- \rangle = &\langle \left [ \Lambda e^{\frac{A }{\sqrt{D} \xi }} -\frac{1}{2}\left(\Pi_A^2 +A_\sigma^2\right) + 2 \sqrt{D}\xi_A  A_{\sigma\sigma} \right ]\rangle +\hbar^2 \frac{D-27}{12} +
\\&+ \langle \left [ \frac{1}{2}\left(\Pi_\emptyset^2 +B_{_\emptyset,\sigma}^2\right) + 2 \sqrt{D}\xi_\emptyset  B_{_\emptyset,\sigma\sigma} \right ]\rangle+\frac{1}{2} \sum_{i=1}^{D-1} \langle \Pi_i^2 +B_{_i,\sigma}^2\rangle =0,
}
\end{equation}
\begin{equation}\fl
\eqalign{
\langle L^+ - L^- \rangle = &\langle \left [ -\Pi_A A_\sigma - 2 \sqrt{D}\xi_A  \Pi_{A,\sigma} \right ]\rangle+
\\&+ \langle \left [ \Pi_\emptyset B_{\emptyset,\sigma} + 2 \sqrt{D}\xi_\emptyset  \Pi_{\emptyset,\sigma} \right ]\rangle+\frac{1}{2} \sum_{i=1}^{D-1} \langle \Pi_\emptyset B_{i,\sigma}\rangle = 0,
}
\end{equation}
\endnumparts
where normal ordering is implied.\\ Solving the first constraint for $\Lambda$ a first result is established: the Liouville field $A$, which is in fact proportional to the conformal mode of the metric (after a conformal transformation), contributes with an opposite sign to the cosmological constant value as compared to the $B$ fields, possibly addressing the cosmological constant problem mentioned in the Introduction: the quantum fluctuations of the dynamical gravitational degrees of freedom, {\it i.e.}, the conformal mode $\varphi \propto A $ in our case, partially compensate the positive contributions to the value of the cosmological constant stemming from the quantum fluctuations of the scalar (matter) fields. This will be shown explicitly for a particular set of states later on.\smallbreak
As a basis for the Hilbert space two possibilities are at hand: coherent states, being eigenstates of the annihilation operators, have the advantage of providing rather simple expressions for the quantum constraints, and therefore seem to be the most obvious choice. On the other hand, since our first goal is to obtain values for $\Lambda$ which follow from the requirement for the lower excitations of the spectrum of the theory to be physical, a Fock basis is the best option.\\
However if the quantum constraints are expressed in terms of creation/annihilation operators the exponential term in $L^{\pm,A}$ would spread every Fock excitation of the field $A$ over the entire spectrum, making the calculation of the matrix elements quite problematic. To avoid this it is possible to use a diagonal representation for the constraint operators in the coherent state (overcomplete) basis \cite{Klauder:1968dq}. This has the advantage of turning all the matrix elements calculations into gaussian integrals over complex variables. By writing a general state as a tensor product of linear combinations of Fock excitations of the Fock vacua we will be able to obtain two constraint equations involving the cosmological constant $\Lambda$.\\
To simplify the picture, we can reorganize the Fock operators used in the quantization of every field:
\begin{equation}\fl
\mathbf{a}_n = \left \{
\begin{array}{l l}
\hat{a}_0 & :\  n = 0,\\ \\
\frac{1}{\sqrt{n}} \hat{a}_n &:\  n > 0,\\ \\
\frac{1}{\sqrt{|n|}} \hat{\bar{a}}_{|n|} &:\   n < 0,\\
\end{array} \right.\quad ,
\qquad 
\mathbf{a}_n^\dagger = \left \{
\begin{array}{l l}
\hat{a}_0^\dagger &\ : n = 0,\\ \\
\frac{1}{\sqrt{n}} \hat{a}_n^\dagger &\ : n > 0,\\ \\
\frac{1}{\sqrt{|n|}} \hat{\bar{a}}_{|n|}^\dagger &\ :  n < 0,\\
\end{array} 
\right. 
\end{equation}
so that $ \left[ \mathbf{a}_n , \mathbf{a}_m^\dagger \right] = \hbar\delta^n_m$.\\
Given the quantum operators $L^\pm(\sigma)$, following the procedure described in \ref{kernelrep}, we get:
\begin{equation} \label{kernelrepconstraints}
\eqalign{
\fl L^\pm(\sigma) &=  \int \prod_m \left[ \frac{d z_m d \bar{z}_m}{2 \pi} \right]  |\ul{z} \rangle \Lslash^\pm (\sigma, z, \bar{z})\langle \ul{z} | =\\
\fl &=  \int \prod_m \left[ \frac{d z_m d \bar{z}_m}{2 \pi} \right]  |\ul{z} \rangle \left ( \Lslash^{\pm,A} (\sigma, z, \bar{z}) + \Lslash^{\pm,\emptyset} (\sigma, z, \bar{z}) + \sum_{i=1}^{D-1} \Lslash^{\pm,i} (\sigma, z, \bar{z}) \right )\langle \ul{z} |,
}
\end{equation}
where $m$ runs over all the modes of creation and annihilation operators. Coherent states are defined as:
\begin{equation}\label{coherentstate}
 | \ul{z} \rangle = \bigotimes_{ X}^{ fields}\left ( \bigotimes_n |z_n^X \rangle \right ),
\end{equation} 
where the first tensor product is over the bosonic fields, excluding the ghosts, by virtue of the BRST symmetry established above. The $\Lslash^{\pm,X}$ are the kernels:
 \numparts
\begin{equation}
\eqalign{
\fl \Lslash^{\pm,A} (\sigma, z, \bar{z}) =&   \frac{\Lambda}{2}  \mbox{exp}\left  [-\frac{1}{4 D\pi \xi^2} \left (1 + 2\sum_{n>0}\frac{1}{n} \right ) \right ] \langle e^{\frac{A }{\sqrt{D} \xi }} \rangle -\\ \fl & - \frac{1}{4}\langle\left(\chi_A^\mp\right)^2 \rangle \mp \sqrt{D} \xi_A  \langle\left( \chi_A^{\mp}\right)_\sigma \rangle + \frac{1}{2\pi}\left (\frac{1}{4}+\sum_{n>0} n \right ) - \hbar^2\frac{D-27}{24},
}
\end{equation}
\begin{eqnarray*}
\Lslash^{\pm,\emptyset} (\sigma, z, \bar{z}) =&  \frac{1}{4}  \langle \left( \chi_0^\pm \right)^2 \rangle \pm \sqrt{D} \xi_\emptyset \langle \left(\chi_0^\pm \right)_\sigma \rangle - \frac{1}{2\pi}\left (\frac{1}{4}+\sum_{n>0} n \right ),\\
\Lslash^{\pm,i} (\sigma, z, \bar{z}) = & \frac{1}{4} \langle \left(\chi_i^\pm\right)^2 \rangle - \frac{1}{2\pi}\left (\frac{1}{4}+\sum_{n>0} n \right ),
\end{eqnarray*}
\endnumparts
where $\langle \mc{O}\rangle$ denotes a diagonal matrix element between two coherent states in the form of \eref{coherentstate}. Once again normal ordering is implied everywhere needed.
The infinite sums\footnote{regulated as specified in \eref{kernel}} appearing in these functions are absorbed in the calculation of matrix elements of \eref{kernelrepconstraints} that will be performed later on.

\paragraph{Physical States in the Fock basis}
As said before, because of the decoupling between the fields of the bosonic sector the Hilbert space is a direct product of the Hilbert spaces for each field. Furthermore each field is described by a mode expansion, so that its Hilbert space is itself a tensor product of independent Hilbert spaces, one for each $n$ labelling the modes. For a single field $X$ a completely general state may be written as:
\begin{equation}\label{physicalstatef}
| \psi^X(d) \rangle = \bigotimes_{n \in \mathbb{Z}} \left [ \sum_{\mu \geq 0} d_\mu^X(n) | \mu_n^X \rangle \right ],
\end{equation}
where $n$ labels the modes, $\mu^X_n$ is the occupation number of the mode $n$ of the field $X$, and the $d$'s are complex coefficients. 
Considering then the whole set of fields in the model, any state in the complete Hilbert space can be written then as:
\begin{equation}\label{physicalstate}
|\psi\rangle = \sum_{\{d^A, d^\emptyset, d^i\}} | \psi^A(d^A) \rangle | \psi^\emptyset(d^\emptyset) \rangle \bigotimes_{i=1}^{D-1}| \psi^i(d^i) \rangle
\end{equation}
where the sum is over an arbitrary number of sets of $d$ coefficients. This choice, which is not the most intuitive\footnote{In the simple example of two decoupled systems, A and B, with an Hilbert space basis $|n\rangle$ and $|m\rangle$ respectively, the easiest way to write a general state has the form $|\psi_g\rangle = \sum_{n,m} \psi(n,m)|n\rangle|m\rangle$. Considering factorized states $|\psi_f (a,b)\rangle = \sum_{n} a(n)|n\rangle\otimes \sum_mb(m)|m\rangle$, a sum over different sets of coefficients $\{a,b\}$ reproduces the general state given the identification $\psi(n,m) = \sum_{a,b} a(n)b(m)$, as in a series expansion.}, has the advantage of providing us with complete control on the single coefficients of every field, so that specific quantum states are easily selected.\smallbreak
To simplify the picture, and take advantage of the decoupling, without loosing insight in the mechanism that constrains the cosmological constant, we will consider a subset of the Hilbert space, in which the quantum states \eref{physicalstate} are defined with a single set of $d$ coefficients, so that the sum is dropped. In this way the quantum state is completely factorized, and we can work in each sector separately.\\
When contracted with coherent states \eref{coherentstate}, it will give, with the condition of unitary norm:
\begin{equation}
\eqalign{
|\langle \ul{z}|\psi\rangle|^2= | \psi (\ul{z}) |^2 = \prod_X | \psi^X (\ul{z}) |^2=\\= \prod_X \prod_n \left [\sum_{\mu, \nu \geq 0} d^X_\mu(n) \bar{d}^{X}_\nu (n) \bar{z}^\mu_n z^\nu_n e^{-|z_n|^2} \right ].
}
\end{equation}
In this way, using the factorization, the constraint equations \eref{quantumconstraints} reduce to a sum of independent integrals over complex variables:
\begin{equation}
\eqalign{
\fl \langle L^\pm(\sigma) \rangle = &\int \prod_m \left[ \frac{d z_m d \bar{z}_m}{2 \pi} \right] \Lslash^{\pm,A} (\sigma, z, \bar{z})  | \psi^X (\ul{z}) |^2+\\ \fl &+ \int \prod_m \left[ \frac{d z_m d \bar{z}_m}{2 \pi} \right] \Lslash^{\pm,\emptyset} (\sigma, z, \bar{z})  | \psi^\emptyset (\ul{z}) |^2 +\\ \fl &+ \sum_i^{D-1} \int \prod_m \left[ \frac{d z_m d \bar{z}_m}{2 \pi} \right]\Lslash^{\pm,i} (\sigma, z, \bar{z})| \psi^X (\ul{z}) |^2.
}
\end{equation}
The integrals are all gaussian in the $z$ variables, since $|\psi|^2$ carries a gaussian factor for each mode. Taking again an orthogonal combination of the constraints, we can finally obtain the equations:
\numparts
\begin{equation}
\eqalign{\label{qc1}
\fl \langle L^+ + L^- \rangle =   \frac{2}{\sqrt{\pi}} \sum'_n |n|^{3/2} \left [ \frac{\sqrt{D}\xi_\emptyset}{\varpi_n^\emptyset}\Im \left (\omega^{(1)\emptyset}_n e^{- i n \sigma}  \right ) - \frac{\sqrt{D}\xi_A}{\varpi_n^A}\Im\left (\omega^{(1)A}_n e^{- i n \sigma}  \right ) \right] - \\
\fl  - \frac{1}{4\pi} \sum_{X}^{\mbox{\tiny fields}} f \Biggl \{ \Bigl ( \sum_{n,m \geq 0} + \sum_{n,m \leq 0} \Bigr )_{n \neq m} \frac{4 \sqrt{n^* m^*}}{\varpi^X_n \varpi^X_m} \Re \left ( \omega^{(1)X}_n e^{- i n \sigma}  \right ) \Re \left ( \omega^{(1)X}_m e^{- i m \sigma}  \right ) +\\
\fl + \Bigl ( \sum'_n + 2\delta^n_0 \Bigr ) \left [ \frac{2 n^*}{\varpi^X_n} \left ( \Re \left ( \omega^{(2)X}_n e^{- i 2 n \sigma}  \right ) + \tilde{\omega}^X_n - 1 \right ) \right ] \Biggr \} - \hbar^2\frac{27-D}{12} + \Lambda \prod_\ell \Upsilon_\ell ,
}
\end{equation}
\begin{equation}
\eqalign{\label{qc2}
\fl \langle L^+ - L^- \rangle = \frac{2}{\sqrt{\pi}} \sum'_n n |n|^{1/2} \left [ \frac{\sqrt{D}\xi_\emptyset}{\varpi_n^\emptyset}\Im \left (\omega^{(1)\emptyset}_n e^{- i n \sigma}  \right ) - \frac{\sqrt{D}\xi_A}{\varpi_n^A}\Im\left (\omega^{(1)A}_n e^{- i n \sigma}  \right ) \right] - \\
\fl  -\frac{1}{4\pi} \sum_{X}^{\mbox{\tiny fields}} f \Biggl \{ \Bigl ( \sum_{n,m \geq 0} - \sum_{n,m \leq 0} \Bigr )_{n \neq m}  \frac{4 \sqrt{n^* m^*}}{\varpi^X_n \varpi^X_m}  \Re \left ( \omega^{(1)X}_n e^{- i n \sigma}  \right ) \Re \left ( \omega^{(1)X}_m e^{- i m \sigma}  \right ) +\\
\fl + \sum'_n \left [ \frac{2 n}{\varpi^X_n} \left ( \Re \left ( \omega^{(2)X}_n e^{- i 2 n \sigma}  \right ) + \tilde{\omega}^X_n  \right ) \right ] \Biggr \} ,
}
\end{equation}
\endnumparts
where $$n^* = \Biggl \{ \begin{array}{c c} \frac{1}{4}&\ : n = 0,\\ |n|&\ : n\neq 0.\end{array}$$
The sum over the fields, with the factor $f$, means that there is one such contribution from each field in the model, with $f=-1$ for the $A$ field and $f=1$ for all the others.$\ \Upsilon_\ell$ is the term coming from the integration of the Liouville term:
\begin{equation}
\eqalign{
\fl \Upsilon_\ell = & \sum_{\mu, \nu \geq 0} \sum_{\alpha = 0}^\mu \sum_{\beta = 0}^\nu \binom{\mu}{\alpha} \binom{\nu}{\beta} d^A_\mu(\ell) \bar{d}^A_\nu(\ell) i^{\mu - \alpha - \nu +\beta} \times \\ \fl & \times \left [ \sum_\gamma^{\alpha+\beta} \binom{\alpha + \beta}{\gamma} \boldsymbol{\mathsf{S}}_\ell^{\alpha + \beta - \gamma} \int_{-\infty }^{+\infty } d Y Y^{\gamma }e^{-Y^2}  \right] \times \\
\fl & \times \left [ \sum_\delta^{\mu + \nu - \alpha- \beta} \binom{\mu + \nu - \alpha- \beta}{\delta} \boldsymbol{\mathsf{C}}_\ell^{\mu + \nu - \alpha- \beta - \delta} \int_{-\infty }^{+\infty } d Y Y^{\delta }e^{-Y^2} \right],
}
\end{equation}
with
\begin{equation}
\boldsymbol{\mathsf{C}}_\ell = \Biggl \{ \begin{array}{l c} 0&\ell = 0\\ \frac{cos(\ell \sigma)}{\sqrt{\pi |\ell|} 2\sqrt{D} \xi}& \ell \neq 0 \end{array}, \qquad \quad
\boldsymbol{\mathsf{S}}_\ell = \Biggl \{ \begin{array}{l c} \frac{i}{2\sqrt{D} \sqrt{\pi} \xi}&\ell = 0\\ \frac{sin(\ell \sigma)}{\sqrt{\pi |\ell|} 2\sqrt{D} \xi}& \ell \neq 0 \end{array},
\end{equation}
and the omega's are combinations of the $d$ coefficients which define the quantum state of the field $X$ they refer to:
\numparts
\begin{eqnarray}\label{omegas}
\varpi^X_n =& \sum_{\mu \geq 0} |d^X_\mu(n) |^2 \mu!,\\
\omega^{(1)X}_n = & \sum_{\mu \geq 0} d^X_\mu(n) \bar{d}^X_{\mu+1}(n)(\mu + 1)!,\\
\omega^{(2)X}_n = & \sum_{\mu \geq 0} d^X_\mu(n) \bar{d}^X_{\mu+2}(n)(\mu + 2)!,\\
\tilde{\omega}^X_n =& \sum_{\mu \geq 0} |d^X_\mu(n) |^2 (\mu+1)!.
\end{eqnarray}
\endnumparts
A second important result is explicit in these equations: while the second of \eref{qc2} provides nothing more than a constraint on the coefficients $d$, the first one may be solved for the cosmological constant, $\Lambda$, for a given physical state, and determines its value as a function of the coupling constant $\xi$ between the scalar degrees of freedom and the Ricci scalar, and the $d$ coefficients themselves. Hence the requirement for a quantum state to be physical, {\it i.e.}, to be annihilated by the quantum constraints, can be realized only for a specific value of $\Lambda$.\\
Furthermore as mentioned before the gravitational and matter sector have opposite contributions to the value of $\Lambda$, as the $f$ factor clearly states.

\section{Spectrum Analysis}
Having obtained the expressions for the quantum constraints in terms of the coefficients which identify quantum states in Hilbert space, in an illustration of what kind of restrictions may arise for $\Lambda$, it is interesting to look into the spectrum of values that the cosmological constant takes when lower excitations of the model are required to be physical states.
\subsection{The Vacuum}
It seems a reasonable assumption for the Fock vacuum of the theory to be a physical state. Moreover in the model we are considering this choice corresponds to a static Minkowski solution.
Since this state is simply the tensor product of all Fock vacua, for every field $X$ and for every mode, the omega's defined in \eref{omegas} will be:
\begin{equation}
\varpi^X_n = \tilde{\omega}^X_n = \Upsilon_n = 1, \quad \omega^{(1)X}_n = \omega^{(2)X}_n = 0,  \quad \forall n \in \mathbb{Z}, \forall X,
\end{equation}
so that \eref{qc2} is identically vanishing, while \eref{qc1} gives:
\begin{equation}
\Lambda = \frac{27 - D}{12} \hbar^2 = \Lambda_\Omega.
\end{equation}
Hence the cosmological constant is forced by the quantum constraints to take a specific value, which is nothing else that the (linear) central charge which appears in the quantum Virasoro algebra \eref{quantumvirasoroextended}. It is then just in an indirect way, {\it i.e.}, via the requirement of a conformal symmetry at the quantum level, that the coupling of scalar degrees of freedom to gravity induces a (generally) non vanishing cosmological constant in the vacuum, in contrast with the classical requirement $\Lambda=0$ for this solution. It is also worth to note that in spite of the dependence on $D$, which relates the cosmological constant to the matter content of the model, $\Lambda_\Omega$ is independent from the coupling constant $\xi$.\\
Such a result is not surprising. In ordinary bosonic string theory the special case $D=26$ ensures a vanishing cosmological constant on the world-sheet, cancelling the anomaly arising from the quantization of the ghosts. In our model of Liouville gravity a dynamical gravitational degree of freedom, namely the Liouville field $A$, is present, providing an additional contribution. 

\subsection{First level excitations}
To go further, we will now show that imposing the same condition on a subset of the first level excitations of the fields\footnote{Namely states with occupation number at most 1 for each field.} will provide a spectrum of values for $\Lambda$, depending on the coefficients which define the quantum state and the coupling constant $\xi$.\\
Considering, for each field $X$, an excited state in the mode $n^X$ and the vacuum in all other modes:
\begin{equation}\fl
|1\rangle = \bigotimes_X^{fields} |1^X \rangle = \bigotimes_X^{fields} \biggl [ \bigotimes_{n \neq n^X}  |\Omega\rangle \biggr ] \otimes \biggl [ d^X_{0} (n^X)| \Omega \rangle + d^X_{1} (n^X) |1^X_{n^X} \rangle \biggr ],
\end{equation}
we can calculate the quantum constraints \eref{qc1} and \eref{qc2} and apply a Fourier transform, so as to eliminate the $\sigma$ dependence. For the zero modes this leads to:
\numparts 
\begin{equation}\label{1stzeromodes+}
\eqalign{
\langle L^+_0 + L^-_0 \rangle &= \hbar^2 \frac{27 -D}{12} - \Lambda \left ( 1 + \frac{|d^X_{1}(n^A)|^2}{4 D \pi \xi^2 |n^A|}\biggr | _{n^A \neq 0} \right ) \\ & + \frac{1}{4 \pi} \sum_X^{\mbox{\tiny fields}} \biggl [ f \left ( 2 |n^X| + \delta^{n^X}_0 \right ) |d^X_{1}(n^X)|^2 \biggr ],
}
\end{equation}
\begin{equation}\label{1stzeromodes-}
\langle L^+_0 - L^-_0 \rangle \propto \sum_X^{\mbox{\tiny fields}} \biggl [ f \ n^X |d^X_{1}(n^X)|^2 \biggr ],
\end{equation}
\endnumparts
while for the other modes:
\numparts
\begin{equation}\label{1stnmodes+}
\langle L^+_n + L^-_{-n} \rangle \propto \Lambda \left (\bar{d}^A_0 (n) d^A_1(n) + d^A_0 (-n) \bar{d}^A_1(-n) \right ),
\end{equation}
\begin{equation}\label{1stnmodes-}
\eqalign{
\langle L^+_n - L^-_{-n} \rangle  = &\xi_\emptyset \biggl (\bar{d}^\emptyset_0 (n) d^\emptyset_1(n) - d^\emptyset_0 (-n) \bar{d}^\emptyset_1(-n) \biggr ) -\\ &- \xi_A \biggl ( \bar{d}^A_0 (n) d^A_1(n) - d^A_0 (-n) \bar{d}^A_1(-n) \biggr ),
}
\end{equation}
\endnumparts
The analysis of such equations is rather complicated with an arbitrary number of scalar fields. We can then consider in detail the simplest (and more strictly constrained) cases, with $D=1$ and $D=2$.
\paragraph{1st excited level with $D=1$ scalar fields} In this case, with only the $A$ and $B_\emptyset$ fields present in the model, the results are quite straightforward: first, the excited fields have to be in a \emph{pure excited level}, {\it i.e.}, $|\psi^X_{n^X}\rangle \propto |1_{n^X} \rangle$, hence $d^X_0(n^X) = 0$. This follows directly from \eref{1stnmodes+} and \eref{1stnmodes-} when we exclude the solution $\Lambda=0$, which makes \eref{1stzeromodes+} inconsistent.\\
The only possible solutions are then:\smallbreak
\begin{enumerate}
\item Both fields are excited\\
$$n^A = n^\emptyset = N,$$
\begin{equation}
\fl \Lambda = \hbar^2\frac{13}{6} \left ( 1 + \frac{1}{4 D \pi \xi^2 |n^A|}\biggr | _{n^A \neq 0} \right )^{-1} = 
\Biggl \{ \begin{array}{l l}
\hbar^2\frac{13}{3}\frac{2 D \pi \xi^2 |N|}{1 + 4 D \pi \xi^2 |N|}&\ : N \neq 0,\\ \\
\hbar^2\frac{13}{6}&\ : N = 0.
\end{array}
\end{equation}
\item Only the $A$ field is excited:
\begin{equation}
n^A = 0: \qquad \qquad \Lambda = \hbar^2\frac{13}{6}-\frac{1}{4\pi}.
\end{equation}
\item Only the $B_\emptyset$ field is excited:
\begin{equation}
n^\emptyset = 0: \qquad \qquad \Lambda = \hbar^2\frac{13}{6}+\frac{1}{4\pi}.
\end{equation}
\end{enumerate}
As one can see there are strict constraints on the values that $n^A$ and $n^\emptyset$ can take: in particular the fields can be excited in non-zero modes only together and in the same mode (which is reminiscent of level-matching conditions in string theory). If on the other hand one of $A$ or $B_\emptyset$ is in its ground state, only the zero mode of the other field can be excited.\\
The spectrum of values that the cosmological constant $\Lambda$ is allowed to take is bounded and discrete, ranging from the minimum between $(\hbar^2\frac{13}{6}-\frac{1}{4\pi})$ and $(\hbar^2\frac{13}{3}\frac{2 D \pi \xi^2 |N|}{1 + 4 D \pi \xi^2 |N|})$ (depending on the value of the coupling constant $\xi$) and $(\hbar^2\frac{13}{6}+\frac{1}{4\pi})$.\\
As one can see from the equations in the case (i) it also becomes infinitely dense to the left of $\Lambda=\hbar^2\frac{13}{6}$ for large $N$.
Furthermore if both fields are excited in their zero modes the cosmological constant will take the same value as obtained in the vacuum, namely $\Lambda_\Omega$.

\paragraph{1st excited level with $D=2$ scalar fields} Increasing the number of scalar fields loosens the restrictions imposed by the constraints. In fact already with two scalar fields Equations \eref{1stzeromodes+}, \eref{1stzeromodes-}, \eref{1stnmodes+} and \eref{1stnmodes-}, when the additional scalar field $B_1$ is excited in its zero mode, do not fix one of the $d$ coefficients, resulting in a $d$-dependent cosmological constant, {\it i.e.}, a finite part of the spectrum of $\Lambda$ is continuous. Of course when only pure excitations are considered, and all the $d$'s are fixed, as is usually done in string theory, the spectrum is discrete. \\
Furthermore the presence of a third field allows for much more freedom for which modes may be excited, removing (or reducing to inequalities) the constraints on the $n^X$'s obtained above. Again the value $\Lambda=0$ is excluded by the quantum constraints. Details about the different cases allowed are in \ref{d=2case}.\\
Summarizing, the cosmological constant takes values from the minimum between $(\hbar^2\frac{25}{12}-\frac{1}{4\pi})$ and $(\hbar^2\frac{25}{3}\frac{D\pi \xi^2 |N|}{1 + 4 D \pi \xi^2 |N|})$ (depending again on the value of the coupling constant $\xi$), and is unbounded from above. It is everywhere discrete except for values within the range between $(\hbar^2\frac{25}{3}\frac{D\pi \xi^2 |N|}{1 + 4 D \pi \xi^2 |N|})$ and $(\hbar^2\frac{25}{12}+\frac{1}{4\pi})$, where countable infinities of the continuous bands mentioned above appear. These bands overlap more or less depending on the value of the coupling constant $\xi$.
Again it is possible to have $\Lambda_1^{(D=2)}=\Lambda_\Omega$, with the necessary, but not sufficient, condition to have an $A$ field excited in its zero mode.\smallbreak
Generalizing the $D=2$ case we can expect the spectrum for $\Lambda$ to consist of an infinite, countable, discrete set of values, in which some continuous bands appear, reflecting the presence of the unconstrained continuous coefficients $d$.

\subsection{Higher excitations}
Due to the highly non-linear form of the quantum constraints \eref{qc1}\eref{qc2} a general analysis of higher excitations of the model is not easy to perform.
Besides the great number of states which would have to be considered, a main issue is given by the $\sigma$-dependence in the Liouville potential term, which in general prevents us from performing a Fourier transform and work with a countable set of quantum constraints.\\
It is anyway worth pointing out, that it is only in higher (non purely) excited states that the quantum corrected coupling constants $\xi_A, \xi_\emptyset$ play a role in determining the value of the cosmological constant, adding more quantum contributions to $\Lambda$ .

\section{Conclusions}
Two main results have been obtained. First, it has been shown that at the quantum level a model of 1+1 dimensional gravity non-minimally coupled to non interacting scalar matter fields (classically equivalent to a specific model of Liouville gravity with minimally coupled scalar matter) uniquely fixes the cosmological constant value, $\Lambda$, inclusive of quantum corrections: in the conformal gauge, the requirement of physical quantum states to be annihilated by the constraint operators, {\it i.e.}, requiring a quantum realization of the conformal symmetry, allows to express $\Lambda$ as a function of the complex parameters determining the quantum states and the coupling constant $\xi$ (including its quantum corrections).\\
When low level excitations of the fields are considered the quantum cosmological constant takes values within a well defined spectrum, which is spread around the reabsorbed central charge of the total Virasoro algebra of the matter and gravitational fields. This is also the value for $\Lambda$ that allows the vacuum to be a physical state. All other values of the cosmological constant are obtained with extra terms (the omegas in \eref{qc1}) and the $\Upsilon$ factor, which depend on the matter content and on the considered quantum state.\\
In those cases that have been analysed, the spectrum is bounded from below, generally discrete, with some continuous bands, located in a well defined and finite region of the real line. Such continuity however is just a direct consequence of the continuum spectrum of states allowed for the fields ({\it i.e.}, the $d$'s in \eref{physicalstate}). In any case, the equations being linear in $\Lambda$, there is only one value of the cosmological constant possible for a given quantum state. On the other hand it may be possible to have different physical states meeting all the required conditions given a value for $\Lambda$, or none at all.\\
Secondly we showed that the gravitational degree of freedom, {\it i.e.} the conformal mode in the metric tensor, yields a negative contribution to the value of the cosmological constant, in contrast with the positive terms coming from the matter sector. This can possibly address the issue of the huge predictions of the cosmological constant value from QFTs, where quantum contributions from the gravitational sector are simply ignored.\\
While the specific choice of the model, which remarkably allows a rather straightforward quantization, does not allow a direct and simple extension of this specific results to 3+1 dimensions and/or to more generic dilaton gravity models, it is important to keep in mind how GDTs arise naturally in dimensional reductions. This strongly suggests that quantum gravity could play a fundamental role in the determination of the cosmological constant, and that a similar properties might arise also in more complicated contexts.

\paragraph{Acknowledgements} SZ benefits from a PhD research grant of the Institut Interuniversitaire des Sciences Nucl\'eaires (IISN, Belgium). This work is supported by the Belgian Federal Office for Scientific, Technical and Cultural Affairs through the Interuniversity Attraction Pole P6/11.

\appendix

\section{Quantization of the bosonic sector}\label{matterquantization}
The mode expansions \eref{Amodes} give:
\begin{equation}
\chi_A^\pm = \frac{1}{\sqrt{\pi}}\left ( \frac{1}{2} ( a_0 + a^\dagger_0 )+ \sum_n' \left ( \begin{array}{c} a_n \\ \bar{a}_n \end{array} \right ) e^{-\varepsilon |n| }e^{\mp i n \sigma} \right ),
\end{equation}
so that one can easily compute:
\begin{eqnarray*}
\chi_A^\pm (\sigma) \chi_A^\pm (\sigma') & = : \chi_A^\pm (\sigma) \chi_A^\pm (\sigma') : + \frac{\hbar}{\pi} \sum_{n>0} n e^{- \varepsilon n } e^{-i n (\sigma - \sigma')}, \\
\left [ \chi_A^\pm (\sigma) \chi_A^\pm (\sigma') \right ] & = \pm i \hbar (\partial_\sigma - \partial_{\sigma'} ) \Delta (\sigma - \sigma') = \pm 2 i \partial_\sigma \Delta (\sigma - \sigma'),
\end{eqnarray*}
where $\Delta$ is the regularized Dirac $\delta$ function,
\begin{equation}
\Delta (\sigma - \sigma') = \frac{1}{2 \pi} \sum_n e^{- \varepsilon |n|} e ^{-i n (\sigma - \sigma')}.
\end{equation}
Handling the sums with care one can compute the commutators:
\begin{eqnarray*}
\eqalign{
 \int d \sigma d \sigma' &\left [ \chi^\mp_A (\sigma)^2  ,  \chi^\mp_A (\sigma')^2 \right ] e ^{\pm i r \sigma} e ^{\pm i s \sigma'} = \\   &=- 4 \hbar (r-s) \int d \sigma   \chi^\mp_A(\sigma)^2  e ^{\pm i (r+s) \sigma} +\\&+ \left ( \frac{2}{3}r^3 + \frac{4}{3}r \right )\hbar^2\delta_{r+s},
}\\ 
 \int d \sigma d \sigma' \left [ \chi^\mp_{A\sigma}  ,  \chi^\mp_{A\sigma'} \right ] e ^{\pm i r \sigma} e ^{\pm i s \sigma'} = - 4 \hbar \pi r^3 \delta_{r+s},\\
\eqalign{
 \int d \sigma d \sigma' &\left ( \left [ \chi^\mp_{A} (\sigma)^2 ,  \chi^\mp_{A\sigma'} \right ] + \left [ \chi^\mp_{A\sigma}  ,  \chi^\mp_A (\sigma')^2 \right ]  \right ) e ^{\pm i r \sigma} e ^{\pm i s \sigma'} = \\ & = - 4 \hbar (r-s) \int d \sigma \chi^\mp_{A\sigma} e ^{\pm i (r+s) \sigma},
}
\end{eqnarray*}
where normal ordering is always implied when needed. Such terms are essentially the same for the $B$ sector.\\
To deal with the exponential term $K(\sigma) = \mbox{exp} (A(\sigma)/ \sqrt{D} \xi )$, which commutes with itself, one first needs to compute:
\begin{eqnarray*}
&\chi^\pm_A (\sigma) K(\sigma') = :\chi^\pm_A (\sigma) K(\sigma') : - \frac{i\hbar}{2 \sqrt{D} \pi \xi} \sum_{n>0} e^{- \varepsilon n} e ^{\mp i n (\sigma - \sigma')} K(\sigma'),\\
&\left [\chi^\pm_A (\sigma), K(\sigma') \right ]= - \frac{i\hbar}{\sqrt{D} \xi} K(\sigma') \Delta (\sigma - \sigma'),
\end{eqnarray*}
which are the building blocks for the terms involved in the algebra:
\begin{eqnarray*}
\eqalign{
\fl &\int d \sigma d \sigma' \left [ \chi^\mp_A (\sigma)^2  ,  K(\sigma') \right ] e ^{\pm i r \sigma} e ^{\pm i s \sigma'} = -\frac{i\hbar}{\xi} \int d \sigma  :\chi^\mp_A (\sigma) K(\sigma) :e ^{\pm i (r+s) \sigma} +  \\
 \fl & +\frac{\hbar^2}{2 D \pi \xi^2} \sum_{n>0} \left ( e^{- 2 \varepsilon ( n + |n+r| )} -e^{- 2 \varepsilon ( n + |n-r| ) } \right ) K_{\pm(r+s)},
}\\
\fl\int\! d \sigma d \sigma' \left ( \left [ \chi^\mp_{A\sigma}  ,  K (\sigma') \right ] + \left [ K (\sigma)  ,  \chi^\mp_{A\sigma'}\right ]  \right ) e ^{\pm i r \sigma} e ^{\pm i s \sigma'} = \mp \frac{\hbar}{\sqrt{D} \xi} (r-s) K_{\pm (r+s)}.
\end{eqnarray*}
At this point we can finally put all terms together and write the quantum algebra for the $A$ sector:
\begin{equation}
\eqalign{
\fl \left [ L_r^{A\pm} , L_s^{A\pm} \right ] =& (r-s)\hbar \Biggl [ \int d \sigma \left ( -\frac{1}{4} : \chi^\mp_A (\sigma)^2 : \mp \sqrt{D}\xi_A \chi^\mp_{A \sigma} \right ) e^{\pm i (r-s) \sigma} +\\ \fl &+\frac{\Lambda}{2}K_{\pm(r+s)} \left (\frac{\xi_A}{\xi}+\frac{\hbar}{8 D \pi \xi^2} \right ) \Biggr ]  +\hbar\left ( \frac{\hbar}{24}r^3 -4 D \pi \xi_A^2 r^3 +\frac{\hbar}{12}r \right ) \delta_{r+s},
}
\end{equation}
and fix the quantum correction in $\xi_A$ to get a Virasoro algebra with a central extension. By choosing $\xi_A = \xi - \frac{\hbar}{8 \sqrt{D} \pi \xi}$
we finally get:
\begin{equation}
\fl \left [ L_r^{\pm,A} , L_s^{\pm,A} \right ] = (r-s)\hbar L^{\pm,A}_{r+s} + \hbar \left [ \left ( \frac{\hbar}{24} - 4 D \pi \left ( \xi - \frac{\hbar}{8 \sqrt{D} \pi \xi} \right )^2 \right ) r^3 + \frac{1}{12}r \right ] \delta_{r+s} .
\end{equation}
With the same procedure one can compute the quantum algebra of the $B_\emptyset$ sector, for which there is no need to fix the quantum correction in order to obtain a Virasoro algebra, since no exponential term is present in this case:
\begin{equation}
\fl \left [ L_r^{\pm,\emptyset} , L_s^{\pm,\emptyset} \right ] = (r-s) \hbar L^{\pm,\emptyset}_{r+s} - \hbar \left [ \left ( \frac{\hbar}{24} - 4 D \pi \xi_\emptyset^2 \right ) r^3 + \frac{\hbar}{12}r \right ] \delta_{r+s} ,
\end{equation}
and the one for the $B_i$ fields:
\begin{equation}
\left [ L_r^{\pm,i} , L_s^{\pm,i} \right ] = (r-s)\hbar L^{\pm, i}_{r+s} - \hbar^2\left ( \frac{1}{24}  r^3 + \frac{1}{12}r \right ) \delta_{r+s} .
\end{equation}

\section{Kernel representation for quantum operators}\label{kernelrep}
Given a countable collection of Fock spaces, labelled with $n \in \mathbb{Z}$, one can build a coherent state as the tensor product:
\begin{equation}
|\ul{z} \rangle = \bigotimes_n |\ul{z}_n \rangle = \prod_n e^{-\frac{1}{2} |z_n|^2} e^{z_n \hat{a}_n^\dagger} |\Omega \rangle ,
\end{equation}
where $|\Omega \rangle$ is the Fock vacuum, while the Fock operators obey the usual commutator $[\hat{a}_n, \hat{a}^\dagger_m] = \delta^n_m$.
The action of the annihilation operators on coherent states gives:
\begin{equation}
\hat{a}_n |\ul{z} \rangle = z_n |\ul{z} \rangle, \qquad \qquad e^{\alpha \hat{a}_n}|\ul{z} \rangle = e^{\alpha z_n}|\ul{z} \rangle.
\end{equation}
For a generic quantum operator $\hat{\mc{O}}$, with $\langle \ul{z} | \hat{\mc{O}} | \ul{z}  \rangle = \mc{O}(z,\bar{z})$, one can define:

\begin{eqnarray*}
\mc{O}(z,\bar{z}) =& \int \prod_m \left[ d^2 y_m e^{i y_m z_m} e^{i \bar{y}_m \bar{z}_m} \right] \tilde{\mc{O}}(y, \bar{y}),\\
\tilde{\mc{O}}(y, \bar{y})=& \int \prod_m \left[ \frac{d^2 z_m}{(2 \pi)^2} e^{-i y_m z_m} e^{-i \bar{y}_m \bar{z}_m} \right] \mc{O}(z,\bar{z}),
\end{eqnarray*}

where $d^2 y = d y d \bar{y}$, and then, considering that 
\begin{equation}\nonumber
e^{-i y_m z_m} e^{-i \bar{y}_m \bar{z}_m} = \langle \ul{z} |e^{-i \bar{y}_m \hat{a}_m^\dagger} e^{-i y_m \hat{a}_m} |\ul{z} \rangle,
\end{equation}
we have:
\begin{equation}
\mc{O}(z,\bar{z}) = \int \langle \ul{z} |\prod_m \left[ d^2 y_m  e^{-i \bar{y}_m \hat{a}_m^\dagger} e^{-i y_m \hat{a}_m} \right] |\ul{z} \rangle \tilde{\mc{O}}(y, \bar{y}).
\end{equation}
In this way the operator can be written as:
\begin{equation}
\hat{\mc{O}} = \int \prod_m \left[ d^2 y_m   e^{-i y_m \hat{a}_m} e^{-i \bar{y}_m \hat{a}_m^\dagger} e^{-y_m \bar{y}_m}\right] \tilde{\mc{O}}(y, \bar{y}),
\end{equation}
where the extra exponential term comes from the swapping of the two exponential operators. By inserting the identity between them:
\begin{equation}
\eqalign{
\hat{\mc{O}} =& \int \prod_m \left[ \frac{d^2 y_m d^2 z_m}{2\pi}  e^{-i y_m z_m} e^{-i \bar{y}_m \bar{z}_m} e^{-y_m \bar{y}_m}\right] \tilde{\mc{O}}(y, \bar{y}) |\ul{z} \rangle\langle \ul{z} | = \\
=& \int \prod_m \left[ \frac{d^2 z_m}{2 \pi} \right]  |\ul{z} \rangle \Oslash (z, \bar{z})\langle \ul{z} |,
}
\end{equation}
where $\Oslash (z, \bar{z})$, the integral kernel of the operator, is expressed as:
\begin{equation}\label{kernel}
\eqalign{
\Oslash (z, \bar{z}) &= \int \prod_m \left[ d^2 y_m  e^{i y_m z_m} e^{i \bar{y}_m \bar{z}_m} e^{-y_m \bar{y}_m}\right] \tilde{\mc{O}}(y, \bar{y}) =\\
&= \mbox{exp}\left( \sum_m\partial_{z_m} \partial_{\bar{z}_m} \right) \mc{O} (z, \bar{z}),
}
\end{equation}
and the sum has to be considered regulated, e.g. with an dampening exponential $e^{-\epsilon|m|}$, with $\epsilon > 0$.

\section{Spectrum analysis with D=2 scalar fields}\label{d=2case}

If all fields are excited we are forced to have pure excitations of the $A$ and $B_\emptyset$ fields, {\it i.e.}, $d_0^A(N) = d_0^\emptyset (N) = 0$:
\begin{enumerate}
\item $n^A = n^\emptyset = N, \qquad n^1 = 0$,
\begin{equation}\label{}\nonumber
\Lambda = \left ( \hbar^2\frac{25}{12} + \frac{|d_1^1(0)|^2}{4 \pi} \right ) \left ( 1 + \frac{1}{4 D \pi \xi^2 |N|}\biggr | _{N \neq 0} \right )^{-1} .
\end{equation}
$\Lambda$ is positive, bounded and continuous.
\item $n^A = -n^\emptyset = N \neq 0, \qquad n^1 \neq 0, \qquad |d^1_1(n^1)|^2 = 2\frac{N}{n^1}$,
\begin{equation}\nonumber
\Lambda = \left ( \hbar^2\frac{25}{12} + \frac{|N|}{\pi} \right ) \left ( 1 + \frac{1}{4 D \pi \xi^2 |N|} \right )^{-1} .
\end{equation}
$\Lambda$ is positive, unbounded and discrete.
\item $ |n^A| \neq |n^\emptyset|, \qquad n^1 \neq 0, \qquad |d^1_1(n^1)|^2 = 2\frac{n^A - n^\emptyset}{n^1}$,
\begin{itemize}
\item $ n^A = 0, \qquad n^\emptyset \neq 0 \qquad \longrightarrow \qquad \Lambda = \hbar^2\frac{25}{12}+\frac{4|n^\emptyset| - 1}{4\pi}$,\\
$\Lambda$ is positive, bounded and discrete.
\item $ n^A \neq 0, \qquad n^\emptyset = 0 \qquad \longrightarrow \qquad \Lambda = \left ( \hbar^2\frac{25}{12} + \frac{1}{4\pi} \right ) \left ( 1 + \frac{1}{4 D \pi \xi^2 |n^A|} \right )^{-1}$,\\
$\Lambda$ is positive, bounded and discrete.
\item $ n^A \neq 0, \qquad n^\emptyset \neq 0  \qquad \longrightarrow $ \\ $\Lambda = \left ( \hbar^2\frac{25}{12} + \frac{1}{2\pi}\left (|n^\emptyset| - |n^A| + | n^\emptyset - n^A|\right )\right ) \left ( 1 + \frac{1}{4 D \pi \xi^2 |n^A|} \right )^{-1}$,\\
$\Lambda$ is positive, unbounded and discrete.
\end{itemize}
\end{enumerate}
\smallbreak
Two fields out of three are excited:
\begin{enumerate}
\item $A$ and $B^1$ excited, $B_\emptyset$ in ground state. $A$ is forced to be purely excited.
\begin{itemize}
\item $n^A = n^1 = 0 \qquad \longrightarrow \qquad \Lambda = \hbar^2\frac{25}{12} + \frac{1}{4\pi} (|d_1^1(0)|^2 - 1 )$,\\
$\Lambda$ is positive, bounded and continuous.
\item $n^A \neq 0, \qquad n^1 \neq 0, \qquad |d_1^1(n^1)|^2 = \hbar^2\frac{n^A}{n^1} \longrightarrow \Lambda = \frac{25}{12} \left ( 1 + \frac{1}{4 D \pi \xi^2 |n^A|} \right )^{-1}$,\\
$\Lambda$ is positive, bounded and discrete.
\end{itemize} 
\item $B_\emptyset$ and $B^1$ excited, $A$ in ground state. $B_\emptyset$  is forced to be purely excited.\\
$n^\emptyset = n^1 = 0 \qquad \longrightarrow \qquad \Lambda = \hbar^2\frac{25}{12} + \frac{1}{4\pi} (1+|d_1^1(0)|^2  )$,\\
$\Lambda$ is positive, bounded and continuous.
\item $A$ and $B_\emptyset$ excited, $B^1$ in ground state. Both fields are forced to be purely excited, while $n^A = n^\emptyset = N$,\\
\begin{equation}\nonumber
\Lambda = \hbar^2\frac{25}{12} \left ( 1 + \frac{1}{4 D \pi \xi^2 |N|}\biggr | _{N \neq 0} \right )^{-1} .
\end{equation}
\end{enumerate}
\smallbreak
Only one field is excited:
\begin{enumerate}
\item Field $A$ purely excited in the zero mode $\longrightarrow \Lambda = \hbar^2\frac{25}{12}-\frac{1}{4\pi}$.\\
\item Field $B_\emptyset$ purely excited in the zero mode $\longrightarrow \Lambda = \hbar^2\frac{25}{12}+\frac{1}{4\pi}$.
\item Field $B^1$ excited in the zero mode $\longrightarrow \Lambda = \hbar^2\frac{25}{12}-\frac{1}{4\pi}|d^1_1(0)|^2$.\\
$\Lambda$ is positive, bounded and continuous.
\end{enumerate}

\section*{References}

\bibliographystyle{unsrt}
\bibliography{references}

\begin{thebibliography}{10}

\bibitem{Weinberg:1989}
Steven Weinberg.
\newblock The cosmological constant problem.
\newblock {\em Rev. Mod. Phys.}, 61(1):1--23, Jan 1989.

\bibitem{Weinberg:2000}
Steven Weinberg.
\newblock {The cosmological constant problems}.
\newblock {\em e-print}, (astro-ph/0005265), 2000.

\bibitem{Carroll:1992}
S.~M. Carroll, W.~H. Press, and E.~L. Turner.
\newblock The cosmological constant.
\newblock {\em Ann. Rev. Astron. Astrophys.}, 30:499--542, 1992.

\bibitem{Rugh:2002}
S.~E. Rugh and H.~Zinkernagel.
\newblock The quantum vacuum and the cosmological constant problem.
\newblock {\em Studies In History and Philosophy of Science Part B: Studies In
  History and Philosophy of Modern Physics}, pages 663--705, 2002.

\bibitem{Efstathiou:1990}
G.~Efstathiou, W.~J. Sutherland, and S.~J. Maddox.
\newblock {The cosmological constant and cold dark matter}.
\newblock {\em Nature}, 348:705--707, 1990.

\bibitem{tegmark:2004}
Max Tegmark and et~al.
\newblock Cosmological parameters from sdss and wmap.
\newblock {\em Phys. Rev. D}, 69(10):103501, May 2004.

\bibitem{Govaerts:2004ba}
Jan Govaerts.
\newblock The cosmological constant of one-dimensional matter coupled quantum
  gravity is quantised.
\newblock {\em Proceedings of the 3rd International Workshop On Contemporary
  Problems In Mathematical Physics (COPROMAPH3), eds. J. Govaerts, M. N.
  Hounkonnou and A. Z. Msezane}, pages 244--272, 2004.

\bibitem{GonzalezDiaz:1987gi}
P.~F. Gonzalez-Diaz.
\newblock {Quantized cosmological constant}.
\newblock {\em Mod. Phys. Lett.}, A2:551--554, 1987.

\bibitem{Pinzul:2005ta}
A.~Pinzul and A.~Stern.
\newblock {Noncommutative AdS(3) with quantized cosmological constant}.
\newblock {\em Class. Quant. Grav.}, 23:1009, 2006.

\bibitem{Fujikawa:1996mk}
Kazuo Fujikawa.
\newblock {A toy model of quantum gravity with a quantized cosmological
  constant}.
\newblock {\em Prog. Theor. Phys.}, 96:863--868, 1996.

\bibitem{Major:1995yz}
Seth Major and Lee Smolin.
\newblock {Quantum deformation of quantum gravity}.
\newblock {\em Nucl. Phys.}, B473:267--290, 1996.

\bibitem{Smolin:1995vq}
Lee Smolin.
\newblock {Linking topological quantum field theory and nonperturbative quantum
  gravity}.
\newblock {\em J. Math. Phys.}, 36:6417--6455, 1995.

\bibitem{Borissov:1996}
Roumen Borissov, Seth Major, and Lee Smolin.
\newblock The geometry of quantum spin networks.
\newblock {\em Class. Quant. Grav.}, 13:3183--3196, 1996.

\bibitem{Gambini:2000ir}
Rodolfo Gambini and Jorge Pullin.
\newblock {Making classical and quantum canonical general relativity computable
  through a power series expansion in the inverse cosmological constant}.
\newblock {\em Phys. Rev. Lett.}, 85:5272--5275, 2000.

\bibitem{Grumiller:2002nm}
D.~Grumiller, W.~Kummer, and D.~V. Vassilevich.
\newblock {Dilaton gravity in two dimensions}.
\newblock {\em Phys. Rept.}, 369:327--430, 2002.

\bibitem{Grumiller:2006rc}
Daniel Grumiller and Rene Meyer.
\newblock {Ramifications of lineland}.
\newblock {\em Turk.J.Phys.}, 30:349--378, 2006.

\bibitem{Nakayama:2004vk}
Yu~Nakayama.
\newblock {Liouville field theory: A Decade after the revolution}.
\newblock {\em Int.J.Mod.Phys.}, A19:2771--2930, 2004.

\bibitem{Jackiw:1982hg}
R.~Jackiw.
\newblock {Liouville field theory: a two-dimensional model for gravity?}
\newblock 1982.
\newblock Dedicated to Bryce DeWitt on occasion of his 60th birthday: To be
  publ. by Adam Hilgar, Bristol.

\bibitem{Teitelboim:1983fg}
Claudio Teitelboim.
\newblock {the Hamiltonian structure of two-dimensional space-time and its
  relation with the conformal anomaly}.
\newblock 1983.
\newblock To appear in anniversary vol. dedicated to Bryce DeWitt on his 60th
  birthday.

\bibitem{Brown:1988}
J.~Brown.
\newblock {\em Lower Dimensional Gravity}.
\newblock World Scientific, 1988.

\bibitem{Berger:1972}
B.~K. Berger, D.~M. Chitre, V.~E. Moncrief, and Y.~Nutku.
\newblock Hamiltonian formulation of spherically symmetric gravitational
  fields.
\newblock {\em Phys. Rev. D}, 5(10):2467--2470, May 1972.

\bibitem{Grumiller:2007wb}
D.~Grumiller and R.~Jackiw.
\newblock {Liouville gravity from Einstein gravity}.
\newblock {\em e-print}, (gr-qc/0712.3775), 2007.

\bibitem{Katanayev1986413}
M.~O. Katanayev and I.~V. Volovich.
\newblock String model with dynamical geometry and torsion.
\newblock {\em Physics Letters B}, 175(4):413 -- 416, 1986.

\bibitem{Katanaev19901}
M.~O. Katanaev and I.~V. Volovich.
\newblock Two-dimensional gravity with dynamical torsion and strings.
\newblock {\em Annals of Physics}, 197(1):1 -- 32, 1990.

\bibitem{Witten:1991yr}
Edward Witten.
\newblock {On string theory and black holes}.
\newblock {\em Phys.Rev.}, D44:314--324, 1991.

\bibitem{Callan:1992zr}
Curtis Callan, Steven Giddings, Jeffrey Harvey, and Andrew Strominger.
\newblock Evanescent black holes.
\newblock {\em Physical Review D}, 45(4):R1005, 1992.

\bibitem{Ikeda:1993aj}
Noriaki Ikeda and K.~I. Izawa.
\newblock {General form of dilaton gravity and nonlinear gauge theory}.
\newblock {\em Prog. Theor. Phys.}, 90:237--246, 1993.

\bibitem{Banks:1991}
T.~Banks and M.~O'Loughlin.
\newblock Two-dimensional quantum gravity in minkowski space.
\newblock {\em Nuclear Physics B}, 362(3):649 -- 664, 1991.

\bibitem{Odintsov:1991}
S.~D. Odintsov and I.~L. Shapiro.
\newblock One loop renormalization of two-dimensional induced quantum gravity.
\newblock {\em Phys. Lett. B}, 263:183--189, 1991.

\bibitem{LouisMartinez:1993eh}
D.~Louis-Martinez, J.~Gegenberg, and G.~Kunstatter.
\newblock {Exact Dirac quantization of all 2-D dilaton gravity theories}.
\newblock {\em Phys. Lett.}, B321:193--198, 1994.

\bibitem{Govaerts:1991}
Jan Govaerts.
\newblock {\em Hamiltonian Quantisation and Constrained Dynamics}, volume~4 of
  {\em Leuven Notes in Mathematical and Theoretical Physics}.
\newblock Leuven University Press, 1991.

\bibitem{Gribov:1977wm}
V.~N. Gribov.
\newblock {Quantization of non-Abelian gauge theories}.
\newblock {\em Nucl. Phys.}, B139:1, 1978.

\bibitem{Difrancesco:1999}
Philippe Di~Francesco, Pierre Mathieu, and David Senechal.
\newblock {\em Conformal Field Theory}.
\newblock Springer, corrected edition, January 1999.

\bibitem{Ginsparg:1988ui}
Paul~H. Ginsparg.
\newblock {Applied conformal field theory}.
\newblock {\em e-print}, (hep-th/9108028), 1988.

\bibitem{Blumenhagen:2009zz}
Ralph Blumenhagen and Erik Plauschinn.
\newblock {Introduction to conformal field theory}.
\newblock {\em Lect. Notes Phys.}, 779:1--256, 2009.

\bibitem{Curtright:1982}
Thomas~L. Curtright and Charles~B. Thorn.
\newblock Conformally invariant quantization of the liouville theory.
\newblock {\em Phys. Rev. Lett.}, 48(19):1309--1313, May 1982.

\bibitem{Kummer:1996hy}
W.~Kummer, H.~Liebl, and D.V. Vassilevich.
\newblock {Exact path integral quantization of generic 2-D dilaton gravity}.
\newblock {\em Nucl.Phys.}, B493:491--502, 1997.

\bibitem{Katanaev:1996ni}
M.O. Katanaev, W.~Kummer, and H.~Liebl.
\newblock {On the completeness of the black hole singularity in 2-d dilaton
  theories}.
\newblock {\em Nucl.Phys.}, B486:353--370, 1997.

\bibitem{Green:1987sp}
Michael~B. Green, J.~H. Schwarz, and Edward Witten.
\newblock {\em {Superstring Theory Vol. 1: Introduction}}.
\newblock Cambridge Monographs On Mathematical Physics. Cambridge Univ. Pr.,
  1987.

\bibitem{Polchinski:1998rq}
J.~Polchinski.
\newblock {\em {String Theory Vol. 1: An Introduction to the Bosonic String}}.
\newblock Cambridge Univ. Pr., 1998.

\bibitem{Klauder:1968dq}
J.~R. Klauder and G.~Sudarshan.
\newblock {\em {Fundamentals of Quantum Optics}}.
\newblock Benjamin, New York, 1968.

\end{thebibliography}

\end{document}